\documentclass[amsmath,amssymb,reprint,twocolumn,superscriptaddress, pra]{revtex4-1}

\usepackage[utf8]{inputenc}
\usepackage[T1]{fontenc}
\usepackage[english]{babel}

\usepackage{amsmath,amssymb,amsfonts}
\usepackage{amstext, mathrsfs, textcomp}

\usepackage{booktabs}
\usepackage{siunitx}

\usepackage{mathptmx}
\usepackage{bigints}  
\usepackage{nicefrac}
\usepackage{multirow}
\usepackage{dcolumn}
\usepackage{bm}

\usepackage{subfigure}
\usepackage{graphicx}
\usepackage{xcolor}
\usepackage[export]{adjustbox}

\usepackage[colorlinks=true, citecolor=blue, urlcolor=blue, linkcolor=blue]{hyperref}

\usepackage{changes}

\graphicspath{
	{Figures/}
}

\newcommand{\TITLE}{Deep learning for nano-photonic materials - The solution to everything!?}

\begin{document}
	\title{\TITLE}
	\author{\firstname{Peter R.} \surname{Wiecha}}
	\email[e-mail~: ]{pwiecha@laas.fr}
	\affiliation{LAAS, Universit\'e de Toulouse, CNRS, Toulouse, France}

\begin{abstract}
Deep learning is currently being hyped as an almost magical tool for solving all kinds of difficult problems that computers have not been able to solve in the past. Particularly in the fields of computer vision and natural language processing, spectacular results have been achieved. The hype has now infiltrated several scientific communities. In (nano-)photonics, researchers are trying to apply deep learning to all kinds of forward and inverse problems. A particularly challenging problem is for instance the rational design of nanophotonic materials and devices.
In this opinion article, I will first discuss the public expectations of deep learning and give an overview of the quite different scales at which actors from industry and research are operating their deep learning models. I then examine the weaknesses and dangers associated with deep learning. Finally, I'll discuss the key strengths that make this new set of statistical methods so attractive, and review a personal selection of opportunities that shouldn't be missed in the current developments.
\end{abstract}

	\maketitle

	In the past year, large language models (LLMs) like ``chat-GPT'', ``GPT-4'' or ``LLM200'' \cite{brownLanguageModelsAre2020, nllbteamNoLanguageLeft2022, bubeckSparksArtificialGeneral2023} and text-to-image generators like ``Midjourney'' or ``stable diffusion'' \cite{rombachHighResolutionImageSynthesis2022, podellSDXLImprovingLatent2023} have demonstrated the literally breathtaking capabilities of large deep learning models.
	As a result, Big Tech companies currently engage in a literal battle over the integration of artificial intelligence (AI) into their products \cite{sanyalBigTechPressure2023}. 
	Supposedly for marketing reasons, the term ``AI'' is thereby often used synonymously for ``deep learning'', which boils down to, a little disrespectfully stated, fitting large mathematical functions in a statistical approach to gigantic amounts of data \cite{goodfellowDeepLearning2016}.
	Although, in contrast to conventional statistical methods, deep learning models are typically much larger in terms of internal degrees of freedom and the number of input / output channels. 
	So large actually, that it becomes extremely hard or even impossible to understand, how a model comes up with its predictions \cite{chakrabortyInterpretabilityDeepLearning2017}.
	Nevertheless, DL has proven to perform almost ``unreasonably effective'' \cite{sejnowskiUnreasonableEffectivenessDeep2020} on problems encompassing an abundant variety of applications.
	Already for more than a decade, deep learning methods are breaking records in computer vision challenges \cite{krizhevskyImageNetClassificationDeep2012, szegedyInceptionv4InceptionResNetImpact2016} or enable previously unimaginable applications in natural language processing \cite{sutskeverSequenceSequenceLearning2014, vaswaniAttentionAllYou2017}.
	Other examples to illustrate the remarkable effectiveness of deep learning, may be protein folding predictions \cite{jumperHighlyAccurateProtein2021}, astrophysical problems like the identification and analysis of galaxy merger events \cite{ferreiraGalaxyMergerRates2020}, or even games like ``Go'', which, before the era of deep learning, was assumed to be unsolvable by computer algorithms  \cite{silverMasteringGameGo2016}.

	\section{Expectations}
	
	In the early stages of applying deep learning to the study of photonic nano-structures and materials, researchers reported successful applications \cite{malkielPlasmonicNanostructureDesign2018, ziatdinovDeepLearningAtomically2017, liuTrainingDeepNeural2018, maDeepLearningEnabledOnDemandDesign2018, wiechaPushingLimitsOptical2019}.
	These findings seemed to validate the prospect of a transformative shift in our computational capacity to address nano-photonics problems.
	In the meanwhile however, several challenges have turned out to be more difficult to overcome than anticipated, and other, unexpected problems have emerged.
	A representative example from the field around nano-photonic materials may be meta-material design.
	For several years already, intense research is taking place with the goal to create deep learning algorithms that increase the accuracy in the conception of photonic metasurfaces  \cite{liuGenerativeModelInverse2018, anDeepLearningApproach2019, wenRobustFreeformMetasurface2020, luPhysicsInformedNeuralNetworks2021, anDeepConvolutionalNeural2021, gahlmannDeepNeuralNetworks2022, gladyshevInverseDesignAlldielectric2023}. 
	However, and despite frequent claims of unprecedented design accuracy, works demonstrating deep learning methods are quite generally at the level of proofs of concept.
	Application oriented and large-scale metasurfaces are still almost exclusively designed via the traditional method based on pre-simulated lookup tables \cite{xuSuperreflectorEnabledNoninterleaved2023, parkAllglass100Mm2023, palermoAllOpticalTunabilityMetalenses2022}.
	As another example, while indeed providing impressive results, also the performance of the initially mentioned protein folding prediction model (``alphaFold2'') has turned out to be less stable and the results less useful, than initially hoped and claimed \cite{jonesImpactAlphaFold2One2022}.
	
	Big Tech has its own examples of overly raised expectations, a prominent one is autonomous driving. 
	Enabling self driving cars by machine learning is in fact an ongoing promise for almost forty years \cite{kanadeAutonomousLandVehicle1986, thorpeAutonomousDrivingCMU1991, mitchellDoesMachineLearning1997, michelsHighSpeedObstacle2005, benglerThreeDecadesDriver2014, grigorescuSurveyDeepLearning2020}.
	Yet, regardless the tremendous research and development efforts over this considerable time-span, progress remains slow, autonomous vehicles still underlie extremely rigid regulations and their market share is insignificant. 
	A recently shut down San Francisco self-driving taxi company employed 1.5 persons per autonomous car, supervising the fleet remotely and intervening every few miles (keep in mind that a conventional taxi requires exactly one driver) \cite{mickleCruiseMovedFast2023}.
	All this despite enormous global invests, in particular during the last decade \cite{higginsSlowSelfDrivingCar2022}.
	
	In summary, the expectations with regards to the power and capabilities of deep learning are tremendous, yet sometimes illusory.
	It is probably safe to claim that these expectations are across the board exaggerated.
	Consequently I believe that it is important to clarify some myths about the capabilities and weaknesses of deep learning methods, and in this light revisit its main limitations, but also strengths and key potentials.

	\section{Yet another deep learning paper?}
	
\begin{figure}
		\centering
		\includegraphics[width=0.75\columnwidth]{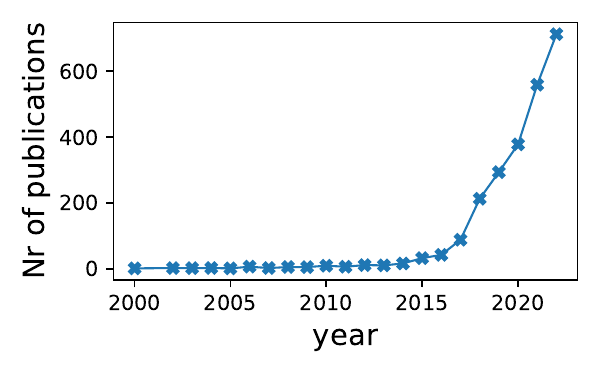}
		\caption{
			Cumulated number of publications per year from 2000 to 2022, combining the topics ``photonics'' and one of: ``deep learning'', ``machine learning'' or ``artificial intelligence''.
			Citation Report graphic is derived from Clarivate Web of Science, Copyright Clarivate 2023. All rights reserved.
		}
		\label{fig:number_publications}
	\end{figure}

	To start, I want to briefly discuss a rarely mentioned problem that is in some ways also a consequence of exaggerated expectations: 
	The flood of publications that can be referred to as ``\textit{yet another deep learning paper}''.
	Because of its medial attention and promises, many researchers try currently to apply deep learning to their problems. 
	Naturally, first tests are often done on simple problems, like design challenges with only a few free parameters, or with very constrained geometries, merely in the perturbation regime. 
	Such problems however, are typically better solved with conventional methods \cite{khaireh-waliehNewcomerGuideDeep2023}. The main reason is the tremendous computational overhead of deep learning. Moreover, being a statistical approach, it also does not make much sense to apply it to problems with only weak variations. 
	However, scientific publication culture has strongly accelerated in the last decades \cite{bornmannGrowthRatesModern2015}. 
	Researchers experience high pressure for publication, especially in an exploding research field, where a year of hesitation can make all the difference. 
	As a consequence, currently an overwhelming number of publications around deep learning applications for photonic materials and nano-optics problems is thrown on the audience (figure~\ref{fig:number_publications}). A significant fraction of these may be cautiously called ``incremental results''.
	On the other hand, there \textit{are} various highly relevant results that merit the community's full attention.
	The ever increasing inflation of publications unfortunately dilutes the important works and renders them more and more difficult to spot. 
	For photonics materials researchers that are new to deep learning and think about applying it to their research, I recommend to get an overview of the more significant results through recent review articles, such as \cite{jiangDeepNeuralNetworks2021, chenArtificialIntelligenceMetaoptics2022, khaireh-waliehNewcomerGuideDeep2023}.

	\section{Considerations about scales}
	
	Another problem associated with overwhelming expectations is the difference between the scale of deep learning in the products of Big Tech companies that we hear about in the mass media, and the scale of deep learning that is accessible to researchers in physics or materials science.

	\paragraph{Big Tech} 
	The threshold of networks with more than a billion parameters has been overcome already many years ago \cite{covingtonDeepNeuralNetworks2016}.
	Today, the challenge that Big Tech has set, are models with trillions of free parameters \cite{fedusSwitchTransformersScaling2022}. 
	Some commercial products have in fact already reached this order of magnitude.
	Google's PaLM for instance has 540 billion parameters \cite{chowdheryPaLMScalingLanguage2022}, and GPT-4 is considered to have almost 2 trillions of trained parameters \cite{yalalovGPT4LeakedDetails2023}.
	Concerning the dataset sizes we face similar orders of magnitude. 
	According to a leaked document, GPT-4 is said to be trained on more than 12 trillion ($10^{13}$!) text samples \cite{yalalovGPT4LeakedDetails2023}, which probably amounts to a significant fraction of the internet's entire accessible information.
	OpenAI's competitors are not significantly behind this data scale. Meta's Llama for example was trained on 2 Trillion text samples \cite{touvronLlamaOpenFoundation2023}.
	
	An incredible amount of computing power is required to configure these massive degrees of freedom and process the vast amount of data. GPT-4 is the current leading example. It has been trained for several months, running in parallel on about 25,000 (25,000!) Nvidia H100 GPUs. 
	A single H100 GPU integrates 80 billion transistors and is capable to perform 2 petaFLOPS ($2\times 10^{15}$ 16 bit floating point operations per second), while running on $700$\,Watts of electric power. 
	This signifies more than 17 Megawatts of continuous power consumption and several tens of Gigawatt hours of electric energy spent for just a single training run.
	According to OpenAI's former and once again CEO Sam Altman, the cost for the training of GPT-4 alone amounts to more than 100 million US dollars \cite{knightOpenAICEOSays2023}.
	Yet, this development is continuing to accelerate. Only a few years back, the second-last generation of large language models (GPT-2, BERT, etc...) required per-training energy in the order of hundreds of Megawatt hours, orders of magnitude less than the latest generation of models \cite{strubellEnergyPolicyConsiderations2019}.
	In the meanwhile, also the hardware costs are immense. As of mid 2023, one H100 GPU alone costs around 40,000\$~\cite{watersHowNvidiaCreated2023}.

	\begin{figure}
	\centering
	\includegraphics[width=0.85\columnwidth]{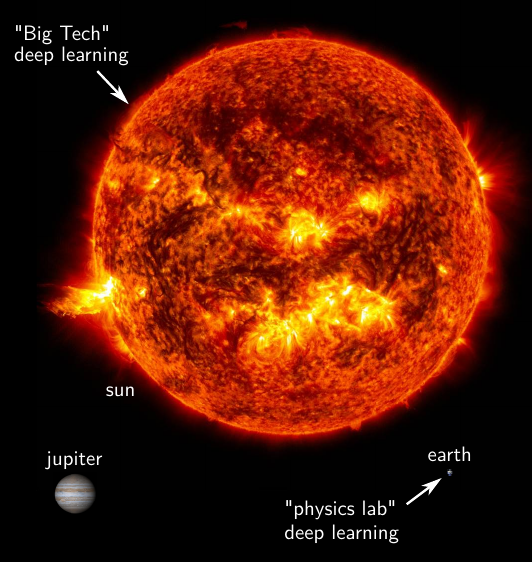}
	\caption{
		Deep learning scale comparison. 
		In almost every metric, deep learning in physics labs is 6 or more orders of magnitude smaller compared to Big Tech applications like ``chat-GPT'' (c.f. table~\ref{tab:NN_size_comparison}). 
		This difference compares like the size of the sun to that of earth, the latter fitting around 1.3 million times in the star at our solar system's center.
		Images by NASA, arranged with permission through NASA's open access policy.
	}
	\label{fig:scale_comparison}
	\end{figure}
	
	\begin{table}[b]
	\caption{
		Orders of magnitude of important metrics in commercial (Big Tech) vs. physics lab scale deep learning.
	}\label{tab:NN_size_comparison}
	\begin{tabular}{lcc} 
		\toprule
		& $\quad$physics lab$\quad$ & Big Tech \\ 
		\midrule
		$N_{\text{gpu}}$      & $1$   & $10^{4}$ \\
		$N_{\text{train time}}$ (GPU hours)  & $10$  & $10^{7}$ \\
		$N_{\text{train cost}}$ (\$) & 100 & $10^{8}$ \\
		$N_{\text{NN params}}$       & $10^7$  & $10^{12}$ \\
		$N_{\text{train samples}}$   & $10^6$  & $10^{13}$ \\
		\bottomrule
	\end{tabular}
	\end{table}

	\paragraph{Physics labs} 
	In comparison to Big Tech, the size of typical deep learning models, their training costs, and the amount of data processed are of a completely different order of magnitude in physics, especially in the fields of nano-optics and photonic devices and materials.
	The number of free parameters are typically in the order of millions to tens of millions ($10^{6}$ - $10^{7}$), the training is often done on single consumer grade GPUs, in general no longer than during a few days.
	In most cases, the amount of available training data is particularly limited, since their generation is usually expensive. Common training sets comprise in the order of ten thousand training samples \cite{wiechaDeepLearningMeets2020,  nohReconfigurableReflectiveMetasurface2022, maBenchmarkingDeepLearningbased2022, khaireh-waliehMonitoringMBESubstrate2023, liuInverseDesignQuantum2023, gostimirovicImprovingFabricationFidelity2023}. 
	Only in rare cases, studies are based on hundreds of thousands ($10^{5}$) or even more samples \cite{maOptoGPTFoundationModel2023, zhangDiffusionProbabilisticModel2023}.
	
	A comparison of the orders of magnitude of the relevant quantities is given in table~\ref{tab:NN_size_comparison}. In fact, Big Tech's neural network models compare to physics lab deep learning in a remarkably consistent way, like comparing the size of our Sun to the size of the planet Earth, as shown at scale in figure~\ref{fig:scale_comparison}. The Earth is about 1.3 million times ($10^{6}$) smaller in volume than the Sun.

	\begin{figure}
		\centering
		\includegraphics[width=0.8\columnwidth]{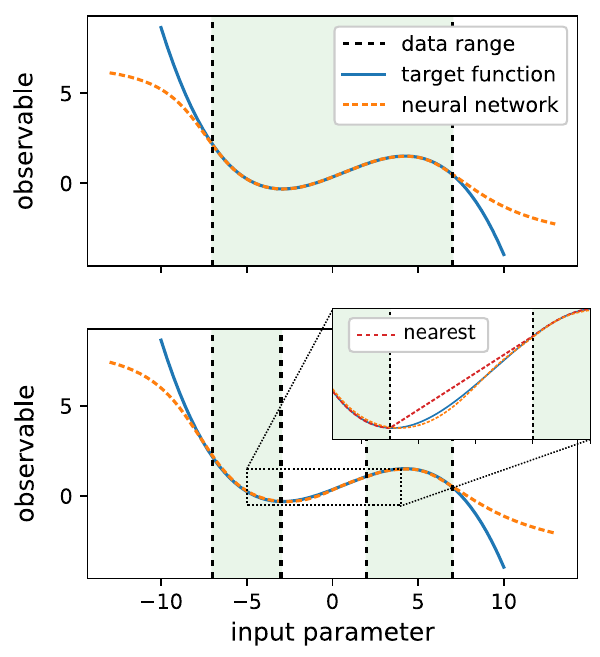}
		\caption{
			Illustration of a neural network's performance at interpolation and extrapolation tasks.
			Two simple 3-layer networks are trained on data, describing a polynomial function. 
			In one case, a larger area of connected data is given (top panel). 
			In a second experiment, the data in a limited parameter region is removed (bottom panel).
			In the interpolation regime where training data is available (inside the green shaded ranges), the accuracy is high. 
			Outside of the data range (left and right sides), extrapolation fails immediately.
			Yet, interpolating between two adjacent zones of available data (white center  area in the bottom panel), works better than the trivial nearest neighbor linear interpolation (red dotted line in the inset).
		}
		\label{fig:interpol_extrapol}
	\end{figure}

	\begin{figure*}
		\centering
		\includegraphics[width=0.9\linewidth]{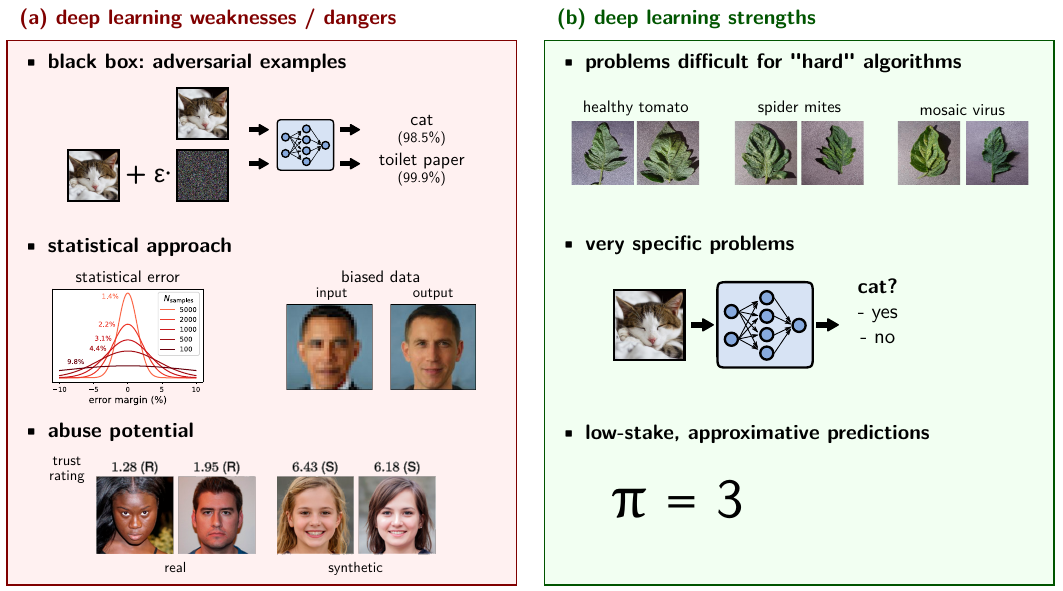}
		\caption{
			Deep learning weaknesses and strengths.
			On the weakness side (a) we have: 
			\textit{Black-box:} Deep learning models are black boxes, untrustworthy and prone to adversarial attacks \cite{goodfellowExplainingHarnessingAdversarial2015}.
			\textit{Statistical approach:} On smaller datasets, deep learning will alwyas add a significant statistical error. For example with 5000 samples, the margin of error for a $2\sigma$ confidence interval is still around $\pm 1.5\%$.
			Furthermore, a neural network will reproduce biases from the datasets, therefore deep learning stands and falls with the data quality.
			\textit{Abuse potential:} Deep learning can be used for generative tasks beyond detectability.
			On the strength side (b) we have: 
			\textit{Hard algorithmic problems:} Problems that are very difficult to solve by conventional algorithms can often be solved with little effort by deep learning. Very detailed image analysis is a good example.
			\textit{Specific problems:} The more specific a problem can be defined, the easier it will be for deep learning.
			\textit{Low stake approximations:} Problems where rough approximations are good enough but speed is all that matters.
			Cat image (c) 2016 by Alexandru Zdrobău, reproduced with permission of the author.
			Obama's white skin upsampling result reproduced from twitter post on 20th June 2020 by user ``chicken3egg''.
			Generated vs real portraits, reproduced with permission from \cite{nightingaleAIsynthesizedFacesAre2022}.
			Tomato leaf dataset images reprinted from \cite{agarwalToLeDTomatoLeaf2020}, copyright 2020, with permission from Elsevier.
		}
		\label{fig:weaknesses_strengths}
	\end{figure*}

	\section{Interpolation and extrapolation}
	
	The main assumption that underlies deep learning is that a mathematical function of large enough degree of freedom can approximate any other function to arbitrary precision \cite{robbinsStochasticApproximationMethod1951}.
	The large mathematical model is the artificial neural network. The target function to be approximated is only implicitly defined by a (large) set of data samples.
	Practically, the parameters of the network are then configured in a fitting process (the training) to match the given set of data. 
	In this process, the hope is that the neural network will develop some kind of global model that describes the data and is able to generalize. However, such generalization happens very rarely, in most cases extrapolation outside the training data region fails quickly \cite{khaireh-waliehNewcomerGuideDeep2023}.
	This is illustrated in figure~\ref{fig:interpol_extrapol} by the example of a multilayer perceptron model, trained on datapoints sampled from a random polynomial function. 
	While interpolation works well, extrapolation fails immediately outside of the training data range.
	
	In nano-optics and photonics materials applications, similar trends are observed. Limiting trained deep learning models to a valid parameter range is therefore an important secondary problem \cite{dengNeuraladjointMethodInverse2021, dengDeepInversePhotonic2022}.
	Especially, because in high-dimensional problems it is often not trivial to specify the interpolation regime \cite{balestrieroLearningHighDimension2021}.
	Figure~\ref{fig:dangers_nanophot}c shows an example of extrapolation to a photonic light router design with around 150 perturbations, whereas the dataset was limited to cases with no more than 20 perturbations \cite{dinsdaleDeepLearningEnabled2021}.
	A possible approach to constrain surrogate models to the interpolation regime of the input space is to use auxiliary networks like variational autoencoders or generative adversarial networks, that learn a compact, regularized representation of the input parameters \cite{melatiMappingGlobalDesign2019, liuHybridStrategyDiscovery2020, augensteinNeuralOperatorBasedSurrogate2023, khaireh-waliehNewcomerGuideDeep2023}.
	It can also help to provide physics through additional loss functions, that test whether a solution is compatible with known physics laws (for example in form of partial differential equations or causality) \cite{raissiPhysicsinformedNeuralNetworks2019, blanchard-dionneTeachingOpticsMachine2020, chenHighSpeedSimulation2022}. 
	But these steps to improve extrapolation require significant efforts, and there is no guarantee to what extent such techniques will work.
	Simply speaking, one must always expect ``to get out what has been put in'', and careful testing of the results is necessary when operating in extrapolation.

	\section{Weaknesses and Dangers}
	
	In addition to the limited extrapolation capabilities discussed before, various further weaknesses and possible pitfalls come with deep learning. 
	A selection is illustrated in figure~\ref{fig:weaknesses_strengths}a. 
	Some limitations can be countervailed with large enough datasets. Others however are inherent to the method. 
	
\begin{figure}
		\centering
		\includegraphics[width=0.9\columnwidth]{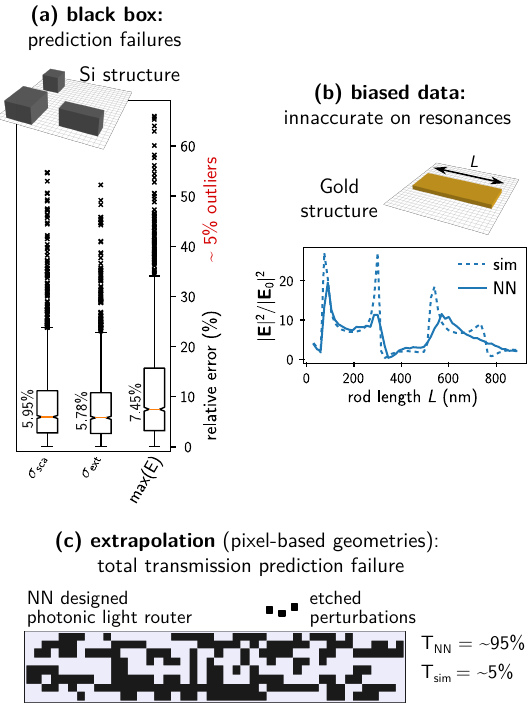}
		\caption{
			Illustrations of the potential impact of deep learning weaknesses on applications in nano-photonics.
			(a) In regression tasks, a certain percentage of predictions is often significantly worse than the statistical average error. Adapted with permission from \cite{wiechaDeepLearningMeets2020}. Copyright 2020 American Chemical Society.
			(b) Infrequent physical phenomena, for example resonances, can easily be under represented in a dataset. Predictions will be inaccurate in such cases. Adapted with permission from \cite{wiechaDeepLearningMeets2020}. Copyright 2020 American Chemical Society.
			(c) Extrapolation to parameter regimes outside the dataset range very generally leads to total prediction failures. Here the network learned photonic structures with up to 20 perturbations that typically feature very high optical transmission. The design network extrapolated in the shown example to a device with $> 150$ perturbations, leading in reality to almost zero transmission of the device. Adapted with permission from \cite{dinsdaleDeepLearningEnabled2021}. Copyright 2021 American Chemical Society.
		}
		\label{fig:dangers_nanophot}
	\end{figure}

	\paragraph{Black Box}
	A fundamental problem of deep learning as such, is the black box character of basically all artificial neural networks.
	In general it is not possible to understand microscopically how a trained neural network makes its predictions. 
	An illustrative symptom of this issue are so-called ``adversarial examples'', that can be found for basically all deep learning image classification models \cite{goodfellowExplainingHarnessingAdversarial2015, suOnePixelAttack2019, liuAdversarialAttacksDefenses2021}. 
	An adversarial example is a weak, but specifically designed noise pattern, that maximally activates a target pathway through the deep learning model and eventually leads to a complete misclassification of the input image.
	Also in various regression models, similar singularities can be found at which predictions totally fail \cite{wiechaDeepLearningMeets2020}. 
	This is illustrated in figure~\ref{fig:dangers_nanophot}a, where, for not understood reasons, a nanophotonics regression model shows prediction failure with very high error in around 5\% of cases.
	This renders neural network predictions untrustworthy in a quite general sense, imposing severe limitations for security-relevant applications such as the above mentioned autonomous driving \cite{qayyumSecuringConnectedAutonomous2020}. 
	For instance, only a few years back it was possible by very simple means to fool Tesla's car assistant system to drive on the wrong side of the street or to ignore speed limits \cite{nassiBAdvertisementAttackingAdvanced2022}.
	Another example for a currently emerging security gap are so-called ``indirect prompt injection'' attacks against large language models like AI-assistants (e.g. Microsoft ``co-pilot'' or Google's ``Bard''). 
	These LLMs are known to be vulnerable to ``engineered'' prompts, which hackers may hide in specifically tailored websites. A model that parses such website and which is equipped with advanced privileges like access to personal data and the internet, can then be forced to execute malicious tasks \cite{greshakeNotWhatYou2023}.
	Despite these known shortcomings, all major Big Tech companies are, as of 2023, about to grant LLM-based personal assistant tools access to the internet and to personal costumer data such as e-mails.
	
	Please note that there are more interpretable machine learning models than deep neural networks \cite{molnarInterpretableMachineLearning2022}, and there are considerable ongoing efforts to render deep learning more explainable \cite{liInterpretableDeepLearning2022, rasExplainableDeepLearning2022}.

	\paragraph{Data quantity and quality}
	Other problems relate to the data quality or quantity, and boil down to the statistical nature of deep learning.
	Naturally, a sufficient amount of data is necessary as the training process will statistically evaluate correlations between the samples. If little data is available, it is therefore important to keep in mind that the predictions of a neural network model will be necessarily worse than the accuracy of the training data. A statistical error will always be added on top.
	This can be circumvented to some degree using data augmentation methods, yet excessive use of such techniques bears the risk of inducing bias in the model \cite{shortenSurveyImageData2019, balestrieroEffectsRegularizationData2022}.
	
	Data quality on the other hand, can pose problems that could also be denominated with the term ``systematic errors''.
	A prominent problem with data quality is biased statistics, that is, data sets that are not representative of the entire parameter space, or that contain an excessive number of samples from a small subspace of the problem. 
	Biased datasets behind natural language processing models or computer vision tools have been widely covered by the media. Both type of applications have reportedly learned racism and sexism, among other social or ethnical biases \cite{ribeiroWhyShouldTrust2016, gebruUsingDeepLearning2017, zouAICanBe2018}.
	In photonics materials design, a typical bias could be the resonant or non-resonant nature of samples in the training data, as illustrated in figure~\ref{fig:dangers_nanophot}b, where the dataset consisted of mostly non-resonant samples \cite{wiechaDeepLearningMeets2020}.
	A neural network may also statistically learn to ignore the possibility of optical losses if trained on a dataset of highly transmissive devices \cite{dinsdaleDeepLearningEnabled2021}.
	Other systematic problems can be caused by erroneous data (e.g. non-converged numerical simulations) or datasets containing many outliers (e.g. noisy measurements) that break the statistical assumptions of the learning process \cite{xiaLearningDiscriminativeReconstructions2015, hendrycksDeepAnomalyDetection2019}.
	Since high quality data is the most essential ingredient for deep learning, in Appendix~\ref{sec:data_quality}, I discuss some of the most frequent problems with data quality and possible solutions.

	\paragraph{Deep fakes}
	Finally, a very dangerous element of deep learning is actually related to the unquestionable strength of large models to generate seemingly real content.
	They genuinely reproduce the characteristics of the training data in new, often undetectable ways. 
	These generative models are nowadays widely used to disseminate deep fakes of images, videos or voices \cite{karrasProgressiveGrowingGANs2017, zakharovFewShotAdversarialLearning2019, karrasAnalyzingImprovingImage2020}.
	In fact, a recent survey that compared real photographs and generated portraits, found that deep learning generated faces were believed to be real significantly more often than photos of actual persons \cite{nightingaleAIsynthesizedFacesAre2022}. 
	The authors of the study assume that the neural network learned to combine key features of faces in a statistically perfectly averaged way, whereas real persons sometimes feature ``outlier'' characteristics (like a slightly crooked nose or asymmetric eyes or ears). Such characteristics of real faces were associated by the test-persons with generation errors of the neural network, whereas the used face generation network was sophisticated at a level to produce seemingly perfect images \cite{karrasAnalyzingImprovingImage2020}.
	As another example, deep voice fakes were reportedly used by criminals to access the bank account of the CEO of a British energy company, relieving him by around a quarter million dollars \cite{stuppFraudstersUsedAI2019}.
	
	Just as waves of AI generated fake news are washed ashore social media platforms \cite{wangThisEraMisinformation2018}, also in science these technologies make it increasingly easier to generate fraudulent results and flood scientific publication channels with fake research \cite{majovskyArtificialIntelligenceCan2023, elaliAIgeneratedResearchPaper2023}.
	This imminent danger can be compared in some ways with distributed denial of service (DDoS) attacks in computer networks. 
	The fraudulent potential can already be impressively illustrated by the two-generations old, openly accessible language model GPT-2 \cite{radfordLanguageModelsAre2019}. It was specifically fine-tuned on arxiv pre-print abstracts, rendering it capable to produce perfectly convincing, yet fake research descriptions. 
	I invite the readers to have a look for themselves on \href{https://thispaperdoesnotexist.netlify.app/}{https://thispaperdoesnotexist.netlify.app/}, keeping in mind that the latest generations of LLMs are incomparably more capable in their reasoning proficiency and processing capacity. Modern models are able to process and generate tens to hundreds of pages of self-consistent, high quality text \cite{bubeckSparksArtificialGeneral2023}.

	\section{Strengths}
	
	Apart from quite a number of risks and dangers that a responsible user should be aware of, there is a reason for the current hype around deep learning. The approach does offer a number of very interesting strengths.
	A selection is illustrated in figure~\ref{fig:weaknesses_strengths}b. 
	
	\paragraph{Problems that are hard for conventional algorithms}
	Deep learning is often very effective at solving problems that are extremely hard to approach with conventional algorithms. In fact, today's AI hype started with an application on such a type of problem: Image classification. Since 2012, no other algorithm could beat deep learning methods in this problem category \cite{krizhevskyImageNetClassificationDeep2012, mahonyDeepLearningVs2019}.
	Countless further examples could be found alone in computer vision. 
	This ranges from classification tasks like tomato plant disease recognition from images of leafs \cite{agarwalToLeDTomatoLeaf2020}, over painting style extraction and its transfer to other images \cite{zhuUnpairedImagetoImageTranslation2017} to entirely generative tasks like bicycle design \cite{regenwetterBIKEDDatasetComputational2021}.

	\paragraph{Very specific problems}
	Similarly, yet to be distinguished from the former point, deep learning performs exceptionally well and is typically very easy to apply on problems that can be defined very specifically.
	This point is particularly true in the case of small and moderate dataset sizes as the amount of required data scales non-linearly with the complexity of a problem. 
	Network models that generalize well usually require gigantic datasets and large efforts in hyper-parameter optimization.
	
	``Fine-tuning'' or ``transfer learning'' are common strategies that exploit the fact that very specific problems are easy to learn. Both of these terms describe similar approaches in which models that have previously been pre-trained on a very large, generic data set are, in a second step, trained on a much smaller data set that contains entirely new samples. The goal is to learn a new, very specific task, either with the same input and output dimensions (fine tuning), or with data of different dimensions that are still of a similar type of problem (transfer learning). 
	This is used for example in computer vision, where pre-trained models learn to interpret images in general \cite{bachmannMultiMAEMultimodalMultitask2022}, and fine-tuning is used subsequently to learn identification of specific, new objects \cite{sladojevicDeepNeuralNetworks2016, kimTransferLearningMedical2022}.
	In natural language processing, fine-tuning is used on models that were pre-trained on a large, yet unspecific data corpus, to learn performing very specific tasks, for instance following explicit, complex instructions, instead of predicting just the next word in a text \cite{wangSelfInstructAligningLanguage2022}.
	In physics and photonics, fine-tuning or transfer learning can be used for example to first teach the general physics to a neural network model through ``cheap'' simulated or analytically calculated data, and fine-tune it subsequently on specific experimental results that are ``expensive'' to generate \cite{ivanovPhysicsbasedDeepNeural2020}. Another application of transfer learning is to migrate concepts between different applications of similar physics, for example from multi-layered planar to multi-shell spherical geometries \cite{quMigratingKnowledgePhysical2019}.

	\paragraph{Low stake, rapid approximations}
	A great strength of deep learning is its capability to deliver rough approximations, fast.
	In time critical applications where stakes are low, this can be an outstanding advantage.
	An everyday application, that many of us have grown used to, are smartphone tools like automatic spelling correcting or next word prediction \cite{hardFederatedLearningMobile2018}.
	It is very useful to get such propositions fast, and a wrongly suggested next word is of no big harm.
	In photonics, equivalent applications can be tools for the rapid estimation of optical performances \cite{nadellDeepLearningAccelerated2019, hegdePhotonicsInverseDesign2020, majorelDeepLearningEnabled2022}, to obtain a rough first suggestion for photonic device designs \cite{estrada-realInverseDesignFlexible2022, haPhysicsdatadrivenIntelligentOptimization2023, luceInvestigationInverseDesign2023},
	or up-scaling techniques for cheap microscopy equipment \cite{rivensonDeepLearningEnhanced2018, sadeghlidizajiMiniaturizedComputationalSpectrometer2022}. 
	But just like the predicted next word on our smartphone app is, such deep learning predictions should always be carefully double-checked.

	\begin{figure}
		\centering
		\includegraphics[width=0.99\columnwidth]{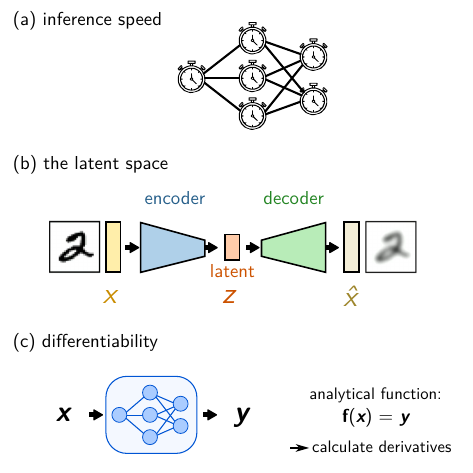}
		\caption{
			The key capabilities and concepts of deep learning in the opinion of the author. 
			(a) the high inference speed of a trained model.
			(b) the concept of the latent space together with various available regularization schemes.
			(c) the analytical character of deep learning models, enabling for example physics informed neural networks or to learn differentiable empirical models even from experimental data.
			Picture of number ``2'' from the MNIST database of handwritten digits \cite{dengMnistDatabaseHandwritten2012}.
		}
		\label{fig:key_strengths}
	\end{figure}
	
	\section{Key capabilities}
	
	Beyond the above list of assets, I personally believe that three specific key attributes of deep learning models, are the most important strengths, that make deep learning stand out with respect to other (statistical) methods. Those points are summarized in figure~\ref{fig:key_strengths}.
	
	\paragraph{Inference speed}
	There's little to add on this fundamental aspect of trained deep learning models: Their evaluation is in general very fast. 
	This is probably the main key strength that is exploited in almost every application.

	\paragraph{The latent space}
	One of the, if not the, most crucial concept in deep learning is the idea of the latent space. 
	A latent description of some thing or concept is a condensed, representative depiction of the object. For instance, the word ``fish'', a fish-symbol or also the according Chinese letter are each in fact some latent description for this kind of animal.
	The appealing capacity of deep learning is, to automatically find latent descriptions via statistical analysis of a large dataset.
	This can even be done on unlabeled, raw data, using unsupervised learning approaches like variational autoencoders (VAEs) or generative adversarial networks (GANs) \cite{goodfellowGenerativeAdversarialNetworks2014, kingmaIntroductionVariationalAutoencoders2019}, often used to tackle inverse problems in photonic materials design \cite{soDesigningNanophotonicStructures2019, jiangFreeFormDiffractiveMetagrating2019}.
	Such approaches are similar to classical principle component analysis \cite{jolliffePrincipalComponentAnalysis2016}, yet the nonlinear, hierarchical character of deep learning models renders them potentially far more powerful.
	Using appropriate regularization techniques, it is possible to extract from the latent space actually meaningful information, like identifying specific properties of persons on portrait pictures (like age, the amount, length or color of hair, the gender, ...) \cite{karrasStyleBasedGeneratorArchitecture2019, abdalImage2StyleGANHowEmbed2019, abdalImage2StyleGANHowEdit2020}.
	The latent space is also the key concept behind generative deep learning. It can be regularized such that every point in a learned latent space corresponds to meaningful generated content or enable smooth interpolation between distant samples in latent space \cite{kingmaIntroductionVariationalAutoencoders2019, sainburgGenerativeAdversarialInterpolative2019, rombachHighResolutionImageSynthesis2022}.

	In physics, learning latent representations can be used for example to identify meaningful coordinate systems. 
	For centuries people believed that the earth is the center of the solar system, yes even of the universe. The debate whether it wouldn't rather be the sun came up, when describing the trajectories of the other planets turned out to be a nightmare in the geocentric coordinate system. With their trajectories forming almost perfect circles in a heliocentric model, the search for adequate coordinate systems has proven to be of crucial importance. 
	Training a network with low dimensional latent space can be helpful to identify such coordinate systems, in which better description of high dimensional, raw observations is possible \cite{championDatadrivenDiscoveryCoordinates2019}, or it allows dimensionality reduction in other problems \cite{kiarashinejadKnowledgeDiscoveryNanophotonics2020, itenDiscoveringPhysicalConcepts2020, zandehshahvarManifoldLearningKnowledge2022}.
	Such reduced representation then often allows an easier identification of correlations in a large dataset.
	One can also include a further, small neural network, into the latent space of a larger mode. This can for instance allow to learn mappings between different latent descriptions, for example to understand system dynamics (by mapping static to dynamic representations) or to discover linear approximations for non-linear physics  phenomena \cite{luschDeepLearningUniversal2018, faselEnsembleSINDyRobustSparse2022}.
	
	A latent space can be learned to categorize data, otherwise difficult to parametrize. In nanophotonics for example, low-dimensional parametrizations can be learned from top-view images of planar nanostructures or of meta-atoms for photonic metasurfaces \cite{liuHybridStrategyDiscovery2020, khaireh-waliehNewcomerGuideDeep2023, augensteinNeuralOperatorBasedSurrogate2023}.
	Furthermore, thanks to the analytical nature of the entire deep learning pipeline (which is a requirement for training via backpropagation), it is possible to include operations like geometric transformations in a model or to work on dynamic discretizations like non isotropic meshes (e.g. from finite element method simulations). Such mesh description can either be learned, or implemented algorithmically, for example via graph neural networks or via hard-coded coordinate transforms \cite{gaoPhyGeoNetPhysicsInformedGeometryAdaptive2020, kuhnExploitingGraphNeural2023, khoramGraphNeuralNetworks2023, deshpandeMAgNETGraphUNet2023}
	
	The latent space does not necessarily need to describe a single property or a single physical object. So-called multi-modal latent models can be built, that map multiple observables, representations or objects into the same latent space \cite{guoDeepMultimodalRepresentation2019}.
	Multi modal physics models can for instance combine different optical properties like far-field and near-field in a single latent space to facilitate according mappings. It could also be possible to combine optical properties of photonic devices with other physical properties like mechanical or thermodynamical attributes. 
	Related approaches have been recently used to infer the optical response in a specific frequency range from data of a different frequency regime \cite{chenCorrelatingMetasurfaceSpectra2023}, or to understand physics across multiple length scales in one model \cite{wanMultiscaleDynamicGraph2019, luMultifidelityDeepNeural2022}.
	
	\begin{figure}
		\centering
		\includegraphics[width=\columnwidth]{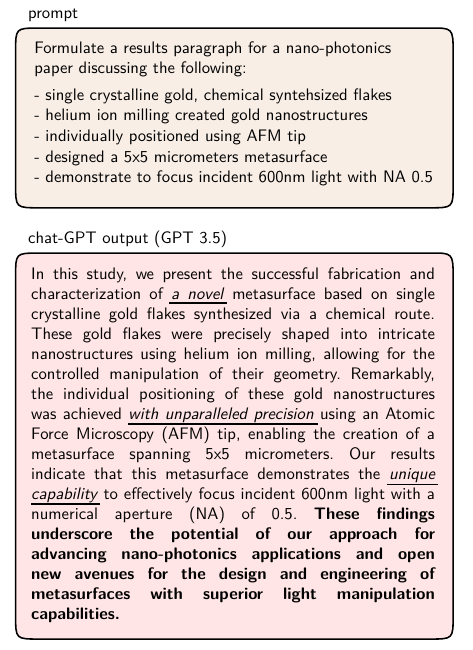}
		\caption{
			Case study to test the capabilities of chat GPT as an editing assistant in scientific research article writing.
			Top: Bullet point list summarizing a fictitious photonics materials research is used as input together with the instruction to formulate a results paragraph for a paper in nano-photonics.
			Bottom: The text generated by chat-GPT (freely accessible version, running on GPT-3.5). 
			Remarkably, chat-GPT inserted novelty claims and exaggerated praising  (italic, underlined) as well as a meaningless babble concluding sentence (bold), not given in the brief summary. 
			Experiment performed in October~2023.
		}
		\label{fig:writing_chatGPT}
	\end{figure}

	\paragraph{Differentiability}
	The third key property of deep learning is the analytic nature of the approach. Remember that an artificial neural network is nothing else than a gigantic mathematical function.
	In consequence, a deep learning model is analytically differentiable.
	This is the key behind so-called physics informend neural networks (PINNs) \cite{raissiPhysicsinformedNeuralNetworks2019, raissiHiddenFluidMechanics2020} and neural operators \cite{liFourierNeuralOperator2021, liPhysicsInformedNeuralOperator2023, kovachkiNeuralOperatorLearning2023}, which aim at learning an approximating to the solution of partial differential equations (PDEs). Using the analytical derivatives of the network predictions, these models can test the validity of a solution within a PDE to arbitrary accuracy, without the need of pre-calculated data.
	In optics, this concept can be used to find approximate solutions to Maxwell's equations \cite{fangDeepPhysicalInformed2020}, the Hemholtz or wave equation \cite{moseleySolvingWaveEquation2020, zhelyeznyakovLargeAreaOptimization2023}, but also for physics-based regularization of data-based models \cite{chenHighSpeedSimulation2022}.
	
	Also aside from PINNs and neural operator networks, the differentiability is a very appealing property. 
	Actually, if enough experimental data is available, but the observations cannot be described with a conventional model, deep learning offers the possibility to train an analytical, empirical model based on actual experimental observation.
	Such differentiable model allows more sophisticated applications than numerical simulations, for example derivatives can help to assess the robustness of solutions against small perturbations \cite{yeungElucidatingBehaviorNanophotonic2020}.
	It also enables gradient based inverse design \cite{renInverseDeepLearning2022, khaireh-waliehNewcomerGuideDeep2023, augensteinNeuralOperatorBasedSurrogate2023}.
	And since the entire calculation pipeline is necessarily analytical and differentiable, it is possible to include other kinds of mathematical operations in the data processing, for example geometry transformations, to render problem descriptions or parametrizations more compact and efficient \cite{gaoPhyGeoNetPhysicsInformedGeometryAdaptive2020}.

	\paragraph{Writing and editorial tools} 
	Last but not least, I believe that aside from direct applications in physics or research on photonic materials, today's deep learning tools can be extremely useful for the daily routine in research. 
	Especially LLMs like chat-GPT can be powerful writing tools, in particular for non-english native speakers, that struggle with the language barrier for the dissemination of their research -- provided they are used in an honest way \cite{stokel-walkerWhatChatGPTGenerative2023}.
	
	An example to illustrate both the potential as well as the dangers associated with such application is shown in figure~\ref{fig:writing_chatGPT}.
	Chat-GPT was instructed to write a paper paragraph given a bullet point list, that describes a fictitious research project about a plasmonic metasurface.
	The generated text is pertinent and contains all provided information in an eloquently written English.
	However, the writing style follows more what would be expected for an abstract and not for a results paragraph. Still, this may be due to the lack of sufficient information. 
	It is far more alarming though, that the model inserted various exaggerated novelty and performance assertions that were not given in the prompt (highlighted by underlining). Finally, chat-GPT even added meaningless prattle to conclude its paragraph (indicated in boldface) -- quite apparently the language model was trained on a corpus of very typical research articles.

%

	\section{Conclusions}
	
	Since the invention of the computer, numerical tools like statistical data processing or physics simulations have enabled unimaginable research insights and opened access to new knowledge \cite{denningComputingDiscipline1989, thijssenComputationalPhysics2007}.
	However, computers and technology in general have accelerated the world drastically. This has a strong impact on all aspects of our lives \cite{turkleLifeScreen2011}.
	Driven by technology and digitalization, also research is accelerating fast, the number of published articles is growing exponentially every year \cite{bornmannGrowthRatesModern2015}.
	Today, deep learning starts to be used in more and more fields of research -- and deep learning is clearly a tool, \textit{designed to accelerate}.
	This development is likely to put additional pressure on every participant in the system, leading to even more self-amplification of these accelerations.
	Moreover, methods of increased efficiency are usually accompanied by a high level of abstraction. This often implies a certain detachment from reality, which in turn carries serious ethical risks \cite{turkleSimulationItsDiscontents2009}.
	
	To conclude I would therefore like to raise some questions:
	What impact will the increased use of accelerating tools like deep learning have on the thoroughness, the integrity and eventually on the quality of research? 
	What will be the impact on the meaningfulness of average scientific publications?
	Will scientists be soon overtaken by the operation speed of their numerical tools?
	Or will big-data driven research rather lead to spectacular new discoveries, comparable to achievements enabled by the era of computer-based simulations?
	
	In order to find positive answers to these these questions, I believe it is advisable to be prudent. Especially regarding the dissemination of effusive claims about the capabilities of deep learning methods in science. 
	I am convinced that deep learning is not the solution to everything.
	It is rather merely a tool, yet unquestionable a powerful one, with tremendous potential for numerous applications.
	Particularly in an age of abundant available data, it bears the potential to reveal hidden correlations, for example through latent space methods, or to allow developing differentiable empirical models from raw observations.
	In the meanwhile it is essential to keep double-checking results, since deep learning models are black boxes, impossible to be categorically trusted. 
	As a last comment, I would like to remind that deep learning cannot create knowledge from nowhere, but relies on hidden information in large datasets. 
	Ideally, its capability to correlate, classify and interpolate intricate data will inspire researchers in their critical thinking and as a result guide the human scientists to the discovery of new phenomena.

	\section*{Acknowledgments}
	
	I thank Arnaud Arbouet, Christian Girard, Otto L. Muskens and Aurélien Cuche for fruitful discussions and the Toulouse HPC CALMIP for their support  through grant p20010. I acknowledge funding from the French Agence Nationale de la Recherche (ANR) under the grants ANR-22-CE24-0002 (project NAINOS) and ANR-22-CE42-0021 (project VERDICT).

	\appendix
	\section{Data quality}\label{sec:data_quality}
	
	As discussed above, high quality data is the foundation of any good deep learning model, and problems with data can have a huge impact on model performance. 
	Furthermore, as data problems stand in the very beginning of the process, they can be very hard to identify in the final model \cite{budachEffectsDataQuality2022}. 
	Therefore I want to briefly discuss common problems with the data and possible approaches to alleviate them:

	\paragraph*{Missing parameter regions / Biased data} If parameter regions are uncovered by the training data, the model will extrapolate in the inference task. This leads to poor, faulty results. 
	Similarly, an over-representation of certain regions in parameter space can lead to reproduction of these biases. 
	In cases where it is hard to estimate if such problems exist, machine learning techniques can be used to assess whether a model extrapolates (e.g. comparing latent space projections of predictions and representative training set samples).
	
	\paragraph*{Redundancies:} If some samples are occurring multiple times, it distorts the statistics, making these samples appear more important than they actually may be. There are more or less sophisticate approaches to identify and remove duplicates, depending for instance on whether duplicates are identical copies or have slightly different signatures \cite{koumarelasDataPreparationDuplicate2020}.
	
	\paragraph*{Errors / Outliers:} Too many such samples break the statistical assumption of the training. Again, latent space embeddings can be used to remove outliers. 
	Data that clusters in latent space are kept, while samples between clusters are removed \cite{fernClusterEnsembleSelection2008, kieuOutlierDetectionMultidimensional2018}. 
	In such cases it is crucial to make sure that the data between clusters are actually outliers and not in fact relevant for the problem.
	Similarity learning or metric learning can help to assess data with difficult statistical properties \cite{belletSurveyMetricLearning2014}.
	
	\paragraph*{Noise:} With noisy data, the predictions may become noisy as well, reducing the accuracy of the results. But in case of random noise, the statistics are still valid, so generating more (noisy) data will improve the situation. Enough training data will then eventually even result in smooth predictions. It may also be possible to apply denoising before training.
	
	\paragraph*{Inconsistent data:} These could be for example incorrect labels, non-converged simulations, a systematic temporal drift in a measurement setup, imperceptible random components of the data source (e.g. in stock market prizes \cite{famaRandomWalksStock1965}), data that partly comes from other generative models \cite{shumailovCurseRecursionTraining2023} etc... 
	It is important to identify such problems since they potentially break the correlations between input and output features. 
	Statistical correlation tests can give a hint of such problems \cite{robinsonCorrelationTestingTime2008}.
	Also machine learning methods such as confident learning exist, that can be used to assess data label quality \cite{northcuttConfidentLearningEstimating2022}.
	
	\paragraph*{Irrelevant information:} Samples that are not related with the actual task can reduce the accuracy on the target problem. This can be particularly problematic with small models and little data. Beginning with a thorough definition of the task can help avoiding such problems and is actually crucial in low-data scenarios.

	Quite generally, a few preventive recommendations can be given that will help avoiding the most frequent problems with data quality.
	
	\begin{itemize}\setlength\itemsep{0cm}
		\item Define the problem well, including a clear data format
		\item Think about data verification early in the process
		\item Perform (conventional) statistical analysis of your data
		\item Verify and document your data sources
	\end{itemize}
	
	Extensive further discussions on data problems and countermeasures can be found in literature \cite{beckerBigDataBig2015}.

	\bibliography{2023_wiecha_deep_learning_solution_to_everything_R1.bbl}

\end{document}